\documentclass[conference]{IEEEtran}
\IEEEoverridecommandlockouts

\usepackage[utf8]{inputenc}
\usepackage[T1]{fontenc}
\usepackage{amsmath,amssymb,amsfonts}
\usepackage{graphicx}
\usepackage{booktabs}
\usepackage{multirow}
\usepackage{cite}
\usepackage{url}
\usepackage{xcolor}
\usepackage{colortbl}
\usepackage{tikz}
\usetikzlibrary{arrows.meta,positioning,calc,fit,patterns,decorations.pathmorphing,backgrounds}

\usepackage{bm}
\usepackage{algorithm}
\usepackage{algpseudocode}
\usepackage{array}

\newcommand{\E}{\mathbb{E}}
\newcommand{\R}{\mathbb{R}}
\newcommand{\KL}{\mathrm{KL}}
\newcommand{\CVaR}{\mathrm{CVaR}}
\newcommand{\relu}{\mathrm{ReLU}}
\DeclareMathOperator*{\argmin}{arg\,min}

\begin{document}

\title{Privacy-Preserving Deep Joint Source-Channel Coding with In-Loop Concept Erasure \thanks{This work is supported by the American University of Beirut University Research Board (URB) and Vertically Integrated Projects (VIP) Program.}}

\author{
\IEEEauthorblockN{Rami Eid, Maria Slim, Mariette Awad, and Hadi Sarieddeen}
\IEEEauthorblockA{\textit{Department of Electrical and Computer Engineering} \\
\textit{American University of Beirut}, Beirut 1107 2020, Lebanon \\
\{rae81,mas194\}@mail.aub.edu; \{ma162, hs139\}@aub.edu.lb}
}
\maketitle

\begin{abstract}

Deep joint source-channel coding (DeepJSCC) transmits learned semantic features efficiently but can leak
sensitive attributes such as gender, race, or speaker identity. We propose LEAPSC (LEACE-in-the-loop privacy for semantic communication), whose core contribution is the integration of in-loop least-squares concept erasure (LEACE) within a variational information bottleneck (VIB) encoder. By periodically refitting the projection operator during training, LEAPSC couples the encoder dynamics to the erasure mechanism, driving attribute-conditional mean differences toward zero within each task-label group on the fitting sample. Additional components, namely
conditional value-at-risk (CVaR) tail-sensitive privacy, feature-wise linear modulation (FiLM)
signal-to-noise ratio conditioning, and Lagrangian dual ascent, improve robustness across channel
conditions and over the high-leakage tail of samples. On CelebA, FairFace, and Google Speech
Commands, LEAPSC reaches task accuracy of $0.862$, $0.755$, and $0.925$ respectively, with attacker accuracy at or below the label-only floor on CelebA ($0.548$ vs.\ floor $0.580$) and within
$2$~percentage points (pp) of chance elsewhere, improving over an information-bottleneck adversarial baseline (IBAL) at a matched $52$-epoch budget by $+3.6$, $+2.5$, and $+1.3$~pp
(Welch's $t$-test, $p{=}0.019$ on CelebA).

\end{abstract}


\section{Introduction}
\label{sec:intro}

Semantic communication prioritizes task-relevant meaning over raw bits~\cite{xie2021deep,bourtsoulatze2019deep}, and deep joint source-channel coding (DeepJSCC)~\cite{bourtsoulatze2019deep,kurka2020deepjscc} has been extended to multi-modal, multi-user, and channel-adaptive regimes~\cite{ismail2026semantic}. Yet the transmitted features typically carry more information than the task requires: an untrusted receiver can train a secondary classifier on the received latents to infer sensitive attributes such as gender, ethnicity, or speaker identity~\cite{li2023privacy,erdogan2023privacy}, a concern magnified in healthcare, voice assistants, and surveillance under General Data Protection Regulation (GDPR) and California Consumer Privacy Act (CCPA) constraints. Deep semantic systems are also vulnerable to adversarial manipulation at the physical layer~\cite{alhaj2026signdeepsc}.

Three families of representation-level privacy coexist: (i) differential privacy~\cite{dwork2006differential}, rigorous but costly in utility on high-dimensional signals; (ii) adversarial training~\cite{edwards2016censoring}, widely used yet unstable and without formal guarantees, since a failed attacker does not preclude a stronger one; (iii) information-theoretic methods~\cite{liao2019information,alemi2017deep}, principled but unable to separate ``irrelevant'' from ``private''. In semantic communications, task-oriented adversarial~\cite{li2023privacy} and rate-distortion~\cite{erdogan2023privacy} formulations follow families (ii)--(iii) and inherit their limitations.

Belrose~et~al.~\cite{belrose2023leace} introduced least-squares concept erasure (LEACE), a closed-form projector that removes all linear predictability of a target concept. It is a strong linear guarantee, but in prior work it has been applied only post-hoc to frozen representations. A trainable encoder can then learn to route private information through directions the fixed projector does not erase, rendering the guarantee vacuous end-to-end. We therefore integrate concept erasure within training, so that representation learning and erasure co-adapt rather than acting in sequence.

The main contribution is an in-loop LEACE mechanism: concept erasure embedded inside the variational information bottleneck (VIB) encoder's training loop with periodic projector refits so that encoder and erasure co-adapt. This yields zero linear leakage on the projector-fitting sample under the LEACE assumptions, at no measurable utility cost. Additional components (conditional value-at-risk (CVaR) tail-sensitive privacy, feature-wise linear modulation (FiLM) signal-to-noise ratio (SNR) conditioning, dual ascent) improve robustness across channel conditions and over the high-leakage tail of samples; robustness is validated via post-hoc and adaptive attackers, multi-seed Welch tests, and comparison against the label-only leakage floor.

\section{System Model and Problem Formulation}
\label{sec:system}

\subsection{Communication System Model}
In our nomenclature, non-bold lowercase letters ($a$) denote scalars, bold lowercase ($\bm{a}$) vectors, and bold uppercase ($\bm{A}$) matrices.
A transmitter (Fig.~\ref{fig:arch}) observes $\bm{x}\!\in\!\mathcal{X}$ (images or audio spectrograms) with task label $y\!\in\!\{1,\ldots,K_y\}$ and sensitive attribute $a\!\in\!\{1,\ldots,K_a\}$, where $K_y$ and $K_a$ denote the number of task and attribute classes, respectively, and encodes $\bm{x}$ into a latent $\bm{z}\!\in\!\R^d$ via $g_\theta$ for transmission over a noisy channel:
\begin{equation}
    \tilde{\bm{z}} = \bm{z} + \bm{n}, \quad \bm{n} \sim \mathcal{N}\!\left(\bm{0},\, \frac{\|\bm{z}\|^2}{d \cdot \mathrm{SNR}} \,\bm{I}_d \right),
    \label{eq:channel}
\end{equation}
where $d$ denotes the latent dimension (set to $128$ in experiments), $\gamma$ denotes the SNR in dB, and $\mathrm{SNR}=10^{\gamma/10}$ is the corresponding linear-scale ratio. The noise variance is normalized to the per-dimension signal power, so the per-coordinate ratio is held at $\mathrm{SNR}$ irrespective of $\|\bm{z}\|$. Because the encoder is conditioned on $\gamma$ via FiLM (Sec.~\ref{sec:film}), the learned mapping depends only on the effective-SNR statistic, so extending to fading requires no architectural change. At the receiver, $f_\phi(\tilde{\bm{z}})$ predicts $y$ while an attacker $\psi(\tilde{\bm{z}})$ attempts to infer $a$. We denote by $\bm{z}$, $\tilde{\bm{z}}$, and $\bm{z}_L$ the encoded, channel-corrupted, and privacy-projected representations, respectively.

\subsection{Threat Model}
We protect the receiver-side privacy-projected representation $\bm{z}_L$, modeling an untrusted downstream consumer of the latent, such as a third-party module the receiver forwards $\bm{z}_L$ to, or an honest-but-curious receiver running its own attribute inference. The adversary has white-box access to $\bm{z}_L$ and may train a nonlinear classifier for $a$, a stronger threat than a linear probe; we evaluate both a fixed-capacity nonlinear probe and an adaptive attacker of higher capacity. We additionally evaluate a black-box attacker that only queries the task predictions $\hat{y}$. An eavesdropper intercepting the raw channel signal $\tilde{\bm{z}}$ is out of scope: that requires channel-level confidentiality, which acts at a different stage than LEAPSC and can be deployed alongside it.

\subsection{Optimization Problem}
\label{sec:opt}
We formulate privacy-preserving semantic transmission as the bi-level constrained program $(\mathcal{P})$ below, comprising an inner attacker problem, an outer defender problem with explicit rate and worst-case privacy constraints, and a Lagrangian relaxation solved by primal-dual ascent (Sec.~\ref{sec:dual}).

Inner problem (attacker): given fixed defender parameters, the attacker maximizes its inference of $a$ from the LEACE-projected $\bm{z}_L = \bm{P}_y \tilde{\bm{z}}$,
\vspace{-1mm}
\begin{equation}
    \psi^* = \argmin_{\psi}\; \E\!\left[H(\psi(\bm{z}_L),\, a)\right],
    \label{eq:attacker}
\end{equation}
where $H(\cdot,\cdot)$ denotes the cross-entropy loss and $\E[\cdot]$ the expectation over the data distribution, denoting the optimal attacker for fixed encoder parameters, approximated by $K_\psi = 3$ stochastic gradient descent (SGD) steps on detached $\bm{z}_L$ per defender update.

Outer problem (defender): the defender (encoder $\theta$, classifier $\phi$, label-conditional prior $\eta$) minimizes task loss subject to a rate constraint and a tail-sensitive privacy constraint, giving the constrained program $(\mathcal{P})$,
\vspace{-1mm}
\begin{equation}
    \min_{\theta,\phi,\eta}\; \E\!\left[\mathcal{L}_\mathrm{task}(f_\phi(\bm{z}_L), y)\right]
    \label{eq:obj}
\end{equation}
\vspace{-5mm}
\begin{align}
    \text{s.t.}\quad & \underbrace{\E\!\left[\KL\!\left(q_\theta(\bm{z}|\bm{x},\gamma) \,\|\, p_\eta(\bm{z}|y)\right)\right]}_{\text{information rate}} \leq \bar{R}(\gamma), \label{eq:rate} \\
    & \underbrace{\CVaR_\alpha\!\left(\ell(\bm{z}_L, a; \psi^*)\right)}_{\text{tail privacy leakage}} \leq \bar{\varepsilon}(\gamma), \label{eq:priv}
\end{align}
where $q_\theta(\bm{z}|\bm{x},\gamma)$ is the encoder posterior conditioned on SNR $\gamma$ via FiLM (Section~\ref{sec:film}), $p_\eta(\bm{z}|y)$ is a learnable label-conditional prior, $\KL(\cdot\,\|\,\cdot)$ denotes the Kullback--Leibler divergence, and the expectation in~\eqref{eq:rate} is taken over the minibatch, matching the implementation, and the budgets $\bar{R}(\gamma)$, $\bar{\varepsilon}(\gamma)$ tighten as channel quality improves.
The privacy proxy $\ell(\bm{z}_L, a; \psi) = \log K_a - H(\psi(\bm{z}_L), a)$ is the per-sample excess cross-entropy above chance; CVaR at $\alpha\!=\!0.9$ focuses the constraint on the most exposed $10\%$ of samples rather than on average leakage.

The overall objective combines task loss with the rate and privacy constraints via Lagrangian relaxation. With dual variables $\beta \geq 0$ (rate) and $\lambda \geq 0$ (privacy),
\vspace{-1mm}
\begin{align}
    \mathcal{L}_\mathrm{def} &= \mathcal{L}_\mathrm{task} + \beta\!\left(\KL - \bar{R}(\gamma)\right) + \lambda\!\left(\CVaR_\alpha(\ell) - \bar{\varepsilon}(\gamma)\right) \notag \\
    &\quad + \lambda_g \mathcal{L}_\mathrm{GRL} + \lambda_m \mathcal{L}_\mathrm{MMD},
    \label{eq:defender}
\end{align}
where $\mathcal{L}_\mathrm{GRL}$ is the loss from a gradient reversal layer (GRL)~\cite{ganin2016domain} and $\mathcal{L}_\mathrm{MMD}$ is the maximum mean discrepancy (MMD) between attribute groups. The $\beta,\lambda$ terms adaptively penalize violations of constraints~\eqref{eq:rate}--\eqref{eq:priv}; the fixed-weight $\lambda_g,\lambda_m$ terms provide complementary regularization.

LEACE zeroes the attribute-conditional mean differences within each task-label group, via projector $\bm{P}_y$ (Section~\ref{sec:leace}). Residual nonlinear leakage is mitigated through adversarial training (the attacker $\psi^*$ pressures the encoder through $\ell$) and CVaR regularization on the high-leakage tail, so a closed-form linear guarantee is complemented by empirical evaluation against nonlinear attackers.

\definecolor{archGray}{RGB}{239,239,239}
\definecolor{archPeach}{RGB}{246,218,202}
\definecolor{archBlue}{RGB}{213,228,241}
\definecolor{archCyan}{RGB}{210,239,243}
\definecolor{archGreen}{RGB}{224,239,211}
\definecolor{archYellow}{RGB}{248,237,194}
\definecolor{archOutline}{RGB}{90,90,90}
\definecolor{archText}{RGB}{55,55,55}

\begin{figure*}[!t]
\centering
\resizebox{0.70\textwidth}{!}{%
\begin{tikzpicture}[
  >=Stealth,
  font=\sffamily,
  arr/.style={-Stealth, line width=1.0pt, draw=archOutline},
  darr/.style={-Stealth, line width=0.8pt, draw=archOutline!65, dashed},
]

\fill[archGray!45, rounded corners=8pt,
      draw=archOutline!22, line width=0.6pt]
  (-0.6,-0.7) rectangle (10.4,5.0);

\fill[archGray!45, rounded corners=8pt,
      draw=archOutline!22, line width=0.6pt]
  (14.0,-2.0) rectangle (24.2,3.4);

\node[font=\normalsize\sffamily\bfseries, text=archText!72]
  at (2.0,4.72) {\textsc{Transmitter}};

\node[font=\normalsize\sffamily\bfseries, text=archText!72]
  at (19.1,3.1) {\textsc{Receiver}};

\fill[archGray, draw=archOutline!65, line width=0.7pt,
      rounded corners=2pt]
  (0,0) rectangle ++(1.8,1.6);

\node[font=\small\sffamily\bfseries, text=archText]
  at (0.9,1.05) {Input};

\node[font=\normalsize\sffamily, text=archText!88]
  at (0.9,0.4) {$\bm{x}$};

\node[font=\footnotesize\sffamily, text=archText!72]
  at (0.9,-0.25) {$64\!\times\!64$};

\fill[archGray, draw=archOutline!65, line width=0.7pt,
      rounded corners=2pt]
  (3.0,-0.2) rectangle ++(2.6,2.0);

\node[font=\small\sffamily\bfseries, text=archText]
  at (4.3,1.15) {ResNet-18};

\node[font=\small\sffamily, text=archText!82]
  at (4.3,0.55) {Backbone $g_\theta$};

\node[font=\footnotesize\sffamily, text=archText!72]
  at (4.3,-0.5) {feat: $512$};

\fill[archCyan, draw=archOutline!65, line width=0.7pt,
      rounded corners=2pt]
  (3.3,2.6) rectangle ++(2.0,0.9);

\node[font=\small\sffamily\bfseries, text=archText]
  at (4.3,3.15) {FiLM};

\node[font=\footnotesize\sffamily, text=archText!82]
  at (4.3,2.8) {$(\bm{s}_f,\bm{t}_f)$};

\fill[archYellow, draw=archOutline!65, line width=0.7pt,
      rounded corners=2pt]
  (3.6,3.9) rectangle ++(1.4,0.7);

\node[font=\small\sffamily\bfseries, text=archText]
  at (4.3,4.25) {SNR $\gamma$};

\fill[archPeach, draw=archOutline!65, line width=0.7pt,
      rounded corners=2pt]
  (6.6,0) rectangle ++(1.6,1.6);

\node[font=\normalsize\sffamily\bfseries, text=archText]
  at (7.4,1.1) {$\bm{\mu}_\theta$};

\node[font=\normalsize\sffamily, text=archText!82]
  at (7.4,0.45) {$\log\!\bm{\sigma}^2_\theta$};

\fill[archPeach!78, draw=archOutline!65, line width=0.7pt,
      rounded corners=5pt]
  (9.15,0.15) rectangle (9.65,1.45);

\node[font=\normalsize\sffamily\bfseries, text=archText]
  at (9.4,0.82) {$\bm{z}$};

\node[font=\footnotesize\sffamily, text=archText!72]
  at (9.4,-0.2) {$d\!=\!128$};

\fill[archBlue, draw=archOutline!65, line width=0.7pt,
      rounded corners=2pt]
  (11.0,0) rectangle ++(2.2,1.6);

\node[font=\small\sffamily\bfseries, text=archText]
  at (12.1,1.1) {AWGN};

\node[font=\small\sffamily, text=archText!82]
  at (12.1,0.5) {Channel};

\fill[archGreen, draw=archOutline!75, line width=0.9pt,
      rounded corners=3pt]
  (14.4,-0.2) rectangle ++(2.8,2.0);

\node[font=\small\sffamily\bfseries, text=archText]
  at (15.8,1.2) {In-Loop};

\node[font=\small\sffamily\bfseries, text=archText]
  at (15.8,0.55) {LEACE};

\node[font=\small\sffamily, text=archText!82]
  at (15.8,-0.0) {$\bm{P}_{\hat{y}}\tilde{\bm{z}}$};

\fill[archGreen!80, draw=archOutline!65, line width=0.7pt,
      rounded corners=5pt]
  (18.15,0.15) rectangle (18.65,1.45);

\node[font=\normalsize\sffamily\bfseries, text=archText]
  at (18.4,0.82) {$\bm{z}_L$};

\fill[archBlue, draw=archOutline!65, line width=0.7pt,
      rounded corners=2pt]
  (20.0,1.5) rectangle ++(2.2,1.3);

\node[font=\small\sffamily\bfseries, text=archText]
  at (21.1,2.4) {Task Classifier};

\node[font=\small\sffamily, text=archText!82]
  at (21.1,1.85) {$f_\phi$};

\fill[archBlue!80, draw=archOutline!65, line width=0.7pt,
      rounded corners=5pt]
  (22.81,1.65) rectangle (23.19,2.65);

\node[font=\normalsize\sffamily\bfseries, text=archText]
  at (23.0,2.15) {$\hat{y}$};

\fill[
  archPeach,
  draw=archOutline!70,
  line width=0.7pt,
  dashed,
  rounded corners=2pt
]
  (20.0,-1.5) rectangle ++(2.2,1.3);

\node[font=\small\sffamily\bfseries, text=archText]
  at (21.1,-0.6) {Attacker};

\node[font=\small\sffamily, text=archText!82]
  at (21.1,-1.05) {$\psi$};

\node[font=\footnotesize\sffamily\itshape, text=archText!72]
  at (21.1,-1.35) {training only};

\fill[
  archPeach!78,
  draw=archOutline!65,
  line width=0.6pt,
  dashed,
  rounded corners=4pt
]
  (22.62,-1.35) rectangle (23.38,-0.45);

\node[font=\normalsize\sffamily\bfseries, text=archText]
  at (23.0,-0.9) {$\hat{a}$};

\node[font=\Large\sffamily\bfseries, text=archText!72]
  at (23.85,-0.9) {\texttimes};

\fill[archBlue!75, draw=archOutline!60, line width=0.7pt,
      rounded corners=2pt]
  (8.5,-2.5) rectangle ++(2.2,1.0);

\node[font=\small\sffamily\bfseries, text=archText]
  at (9.6,-1.95) {Prior $p_\eta(\bm{z}|y)$};

\fill[archYellow, draw=archOutline!65, line width=0.8pt,
      rounded corners=2pt]
  (12.0,-2.5) rectangle ++(2.2,1.0);

\node[font=\small\sffamily\bfseries, text=archText]
  at (13.1,-1.9) {Dual Ascent};

\node[font=\footnotesize\sffamily, text=archText!78]
  at (13.1,-2.2) {$\beta,\lambda$};

\fill[archGreen!75, draw=archOutline!60, line width=0.7pt,
      rounded corners=2pt]
  (15.5,-2.5) rectangle ++(1.6,1.0);

\node[font=\small\sffamily\bfseries, text=archText]
  at (16.3,-1.95) {Label $y$};

\fill[archPeach!75, draw=archOutline!60, line width=0.7pt,
      rounded corners=2pt]
  (12.2,-4.0) rectangle ++(1.8,0.8);

\node[font=\small\sffamily\bfseries, text=archText]
  at (13.1,-3.55) {CVaR$_\alpha$};

\draw[arr] (2.07,0.8) -- (3.0,0.8);
\draw[arr] (5.87,0.8) -- (6.6,0.8);
\draw[arr] (8.30,0.8) -- (9.05,0.8);
\draw[arr] (9.9,0.82) -- (11.0,0.8);
\draw[arr] (13.47,0.8) -- (14.4,0.8);
\draw[arr] (17.55,0.8) -- (17.9,0.82);

\draw[arr]
  (18.9,0.82) --
  (19.5,0.82) --
  (19.5,2.15) --
  (20.0,2.15);

\draw[arr] (22.47,2.15) -- (22.62,2.15);

\draw[darr]
  (19.5,0.82) --
  (19.5,-0.85) --
  (20.0,-0.85);

\draw[darr]
  (22.47,-0.85) --
  (22.62,-0.85);

\node[font=\footnotesize\sffamily, text=archText!72]
  at (10.45,1.1) {$\bm{z}$};

\node[font=\footnotesize\sffamily, text=archText!72]
  at (13.9,1.1) {$\tilde{\bm{z}}$};

\draw[arr, draw=archOutline!70]
  (4.3,3.9) -- (4.3,3.5);

\draw[arr, draw=archOutline!70]
  (4.3,2.6) -- (4.3,1.8);

\node[
  font=\small\sffamily\bfseries,
  text=archText!78
] at (5.35,2.20) {SNR conditioning};

\draw[darr, draw=archOutline!68]
  (20.0,2.6) -- (15.8,2.6) -- (15.8,1.8);

\node[font=\footnotesize\sffamily, text=archText!72]
  at (17.9,2.85) {$\hat{y}$ gating};

\draw[darr, draw=archOutline!68]
  (9.6,-1.5) -- (9.6,0.15);

\node[font=\footnotesize\sffamily\bfseries, text=archText!75]
  at (10.05,-0.45) {KL};

\draw[darr, draw=archOutline!68]
  (16.3,-1.5) -- (16.3,-0.2);

\node[font=\footnotesize\sffamily\itshape, text=archText!72]
  at (17.0,-0.95) {refit};

\draw[darr, draw=archOutline!58]
  (15.5,-2.0) -- (10.7,-2.0);

\draw[-Stealth, line width=0.95pt, draw=archOutline!78]
  (13.1,-3.2) -- (13.1,-2.5);

\draw[darr, draw=archOutline!58]
  (21.1,-1.5) --
  (21.1,-3.55) --
  (14.0,-3.55);

\draw[
  -Stealth,
  line width=0.95pt,
  draw=archOutline!75,
  dashed
]
  (12.55,-1.5)
  -- (12.55,-1.05)
  -- (9.9,-1.05);

\draw[
  -Stealth,
  line width=0.95pt,
  draw=archOutline!75,
  dashed
]
  (13.65,-1.5)
  -- (13.65,-0.82)
  -- (16.8,-0.82)
  -- (16.8,-0.15);

\node[
  font=\small\sffamily\itshape,
  text=archText!72
] at (11.25,-0.83)
  {$R\!\leq\!\bar{R}(\gamma)$};

\node[
  font=\small\sffamily\itshape,
  text=archText!72
] at (17.85,-0.50)
  {$C\!\leq\!\bar{\varepsilon}(\gamma)$};

\end{tikzpicture}%
}

\caption{LEAPSC architecture. Main path (solid arrows): $\bm{x}$ is encoded by a ResNet-18 backbone with FiLM conditioning on SNR $\gamma$, reparameterized into $\bm{z}\!\sim\!q_\theta(\bm{z}|\bm{x},\gamma)$ with $d\!=\!128$, and sent over an AWGN channel producing $\tilde{\bm{z}}$. The receiver computes a gating prediction $\hat{y}=f_\phi(\tilde{\bm{z}})$, selects the in-loop LEACE projector $\bm{P}_{\hat{y}}$ to form $\bm{z}_L$, and re-applies $f_\phi$ to $\bm{z}_L$ for the task prediction; the true label $y$ only groups samples when the projectors are refit every $T_r$ epochs (dashed). The attacker $\psi$ (dashed, training only) infers $\hat{a}$. The label-conditional prior provides the KL rate regularizer, the multipliers $\beta,\lambda$ penalize budget violations via subgradient ascent, and the audited quantity $C$ is attacker accuracy in excess of the label-only baseline.}
\label{fig:arch}
\end{figure*}
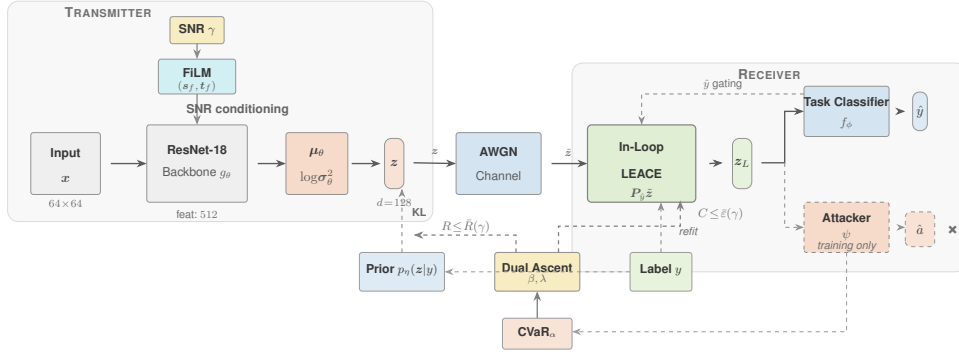

\section{Proposed Methodology}
\label{sec:method}

Each component of LEAPSC maps to a term in problem $(\mathcal{P})$: the VIB encoder (Sec.~\ref{sec:vib}) controls the rate constraint~\eqref{eq:rate}, in-loop LEACE (Sec.~\ref{sec:leace}) provides linear concept erasure, FiLM (Sec.~\ref{sec:film}) makes both SNR-aware, CVaR (Sec.~\ref{sec:cvar}) focuses privacy on the high-leakage tail, and dual ascent (Sec.~\ref{sec:dual}) closes the bi-level solver.

\subsection{Variational Information Bottleneck}\label{sec:vib}
The encoder $g_\theta$ parameterizes a Gaussian posterior $q_\theta(\bm{z}|\bm{x},\gamma) = \mathcal{N}(\bm{\mu}_\theta(\bm{x},\gamma), \mathrm{diag}(\bm{\sigma}^2_\theta(\bm{x},\gamma)))$ over the $d$-dimensional latent space, conditioned on the SNR $\gamma$ through FiLM for consistency with~\eqref{eq:rate}.
Sampling employs the reparameterization trick: $\bm{z} = \bm{\mu}_\theta + \bm{\sigma}_\theta \odot \bm{\epsilon}$, $\bm{\epsilon} \sim \mathcal{N}(\bm{0}, \bm{I})$~\cite{kingma2014auto}, where $\odot$ denotes element-wise multiplication.

To encourage class-specific structure, we adopt a learnable label-conditional prior $p_\eta(\bm{z}|y) = \mathcal{N}(\bm{\mu}_\eta(y), \mathrm{diag}(\bm{\sigma}^2_\eta(y)))$ with per-class embedding parameters $\eta = \{\bm{\mu}_\eta^{(c)}, \log\bm{\sigma}_\eta^{(c)}\}_{c=1}^{K_y}$, where $c$ indexes the $K_y$ label values.
The KL divergence between the two diagonal Gaussians admits a closed-form expression and provides a tractable information-rate regularizer.

\subsection{FiLM-Based SNR Conditioning}\label{sec:film}
To let the encoder adapt its latent structure to channel quality without retraining, we apply feature-wise linear modulation (FiLM)~\cite{perez2018film} to the encoder features. A small MLP maps the normalized SNR $\gamma/10$ to scale and shift vectors:
\begin{equation}
    (\bm{s}_f, \bm{t}_f) = \mathrm{MLP}_\mathrm{FiLM}(\gamma/10), \quad
    \bm{h}' = (\bm{1} + \bm{s}_f) \odot \bm{h} + \bm{t}_f,
    \label{eq:film}
\end{equation}
where $\bm{h} \in \R^{512}$ is the backbone feature vector, $\bm{h}'$ the modulated feature vector, $\bm{s}_f, \bm{t}_f \in \R^{512}$ the predicted scale and shift vectors, and $\bm{1}$ the all-ones vector.

\subsection{In-Loop LEACE}
\label{sec:leace}
A frozen LEACE projector erases the directions predictive of $a$ at the moment of fitting; as the encoder trains, it can subsequently route private information into directions outside the fixed null space, leaving the linear guarantee intact geometrically but vacuous operationally on the final encoder. Refitting the projector inside the training loop makes the erasure follow the representation as it changes, rather than being fixed at a single point in training.

LEACE~\cite{belrose2023leace} constructs a projector $\bm{P}_y$ that zeroes out the directions from which $a$ is linearly predictable on the fitting sample, conditioned on $y$.
A separate projector is fitted for each of the $K_y$ label values:
\begin{equation}
    \bm{P}_y = \bm{I}_d - \hat{\bm{\Sigma}}_y^{1/2}\, \bm{U}_y \bm{U}_y^\top \hat{\bm{\Sigma}}_y^{-1/2},
    \label{eq:leace}
\end{equation}
where $\hat{\bm{\Sigma}}_y$ is the shrunk covariance of $\bm{\mu}_\theta$ within label group $y$ (shrinkage coefficient $5\times10^{-3}$), and $\bm{U}_y$ contains the left singular vectors of $\hat{\bm{\Sigma}}_y^{-1/2} \bm{M}_y$ corresponding to singular values exceeding $10^{-6}$.
Here, $\bm{M}_y$ is the matrix of attribute-conditional mean differences within group $y$.
By construction, $\bm{P}_y \bm{M}_y = \bm{0}$, removing the linearly predictive concept directions identified by LEACE within each fitted label group.

Unlike post-hoc LEACE applied to frozen representations, we refit the projectors every $T_r = 5$ epochs using a rolling buffer of $(\bm{\mu}_\theta, a, y)$ triples collected during training, forcing encoder and projector to co-adapt. Section~\ref{sec:results} quantifies the resulting gain over post-hoc erasure on the same backbone and matched seeds.

At inference $y$ is unavailable, so the receiver computes a gating prediction $\hat{y} = f_\phi(\tilde{\bm{z}})$ on the unprojected output, used only to select the projector, and forms $\bm{z}_L = \bm{P}_{\hat{y}}\,\tilde{\bm{z}}$. The task prediction and the loss in~\eqref{eq:obj} use $f_\phi(\bm{z}_L)$; $\bm{z}_L$ is what any downstream consumer sees. The $K_y$ projectors from the final in-loop refit are shipped with the model, so selection is a constant-time index lookup, and on gating mistakes the resulting attribute accuracy remains near chance.

\subsection{CVaR Tail-Sensitive Privacy}\label{sec:cvar}
Standard average-case privacy constraints may leave highly exposed samples underprotected.
We employ the CVaR~\cite{rockafellar2000optimization} of the per-sample privacy proxy $\ell(\bm{z}_L, a) = \log K_a - H(\psi(\bm{z}_L), a)$ at confidence level $\alpha = 0.9$:
\begin{equation}
    \CVaR_\alpha(\ell) = \min_{\tau}\left\{\tau + \frac{1}{1-\alpha}\,\E\!\left[\relu(\ell - \tau)\right]\right\},
    \label{eq:cvar}
\end{equation}
where $\tau$ is a learnable threshold optimized jointly with the model.
This focuses the privacy constraint on the upper $10\%$ high-leakage tail of samples rather than only average leakage.

\subsection{Dual Ascent and Budget Schedules}\label{sec:dual}
The Lagrangian~\eqref{eq:defender} is solved by alternating primal updates of the model parameters with dual updates of the constraint multipliers. Following the bi-level formulation of Sec.~\ref{sec:opt}, each iteration first performs $K_\psi = 3$ inner SGD steps on the attackers using detached $\bm{z}_L$ to approximate $\psi^*$, after which the defender minimizes $\mathcal{L}_\mathrm{def}$ with respect to $(\theta,\phi,\eta)$.
The multipliers $\beta$ and $\lambda$ are updated via projected subgradient ascent with Robbins-Monro decaying step sizes:
\begin{align}
    \beta_{t+1} &= \left[\beta_t + \frac{\rho_0}{1+t/T_d}\!\left(R_t - \bar{R}(\gamma)\right)\right]_{\!+}, \label{eq:dual_beta}\\
    \lambda_{t+1} &= \left[\lambda_t + \frac{\rho_0}{1+t/T_d}\!\left(C_t - \bar{\varepsilon}(\gamma)\right)\right]_{\!+}, \label{eq:dual_lam}
\end{align}
where $R_t$ and $C_t$ are the audited rate and privacy leakage on a held-out set, $\rho_0 = 0.05$ is the initial step size (denoted $\rho$ rather than $\eta$ to avoid collision with the prior parameters $\eta$), and $T_d = 20$.
The audited $C_t$ is the attacker accuracy in excess of the label-only baseline rather than the CVaR quantity in~\eqref{eq:priv}, so $\lambda$ is an empirical adaptive privacy multiplier driven by an audited leakage surrogate, not an exact dual variable for the CVaR constraint. Being non-differentiable, $C_t$ enters only the subgradient step and requires no backpropagation through the attacker.
The SNR-dependent budgets use sigmoid schedules:
$\bar{R}(\gamma) = (R_{\min} + (R_{\max}-R_{\min})\sigma(-(\gamma-\gamma_0)/s)) \cdot d$ and
$\bar{\varepsilon}(\gamma) = \varepsilon_{\min} + (\varepsilon_{\max}-\varepsilon_{\min})\sigma(-(\gamma-\gamma_0)/s)$,
where $\sigma(\cdot)$ is the logistic sigmoid, $\gamma_0 = 15$~dB the midpoint of the training SNR range, and $s = 4$ the transition steepness. The per-dimension rate endpoints are $R_{\min} = 2$ and $R_{\max} = 6$ nats, scaled by $d$ to match the total KL, and the privacy endpoints are $\varepsilon_{\min} = 0.05$ and $\varepsilon_{\max} = 0.30$.
These schedules allow tighter privacy at higher SNR, where the channel contributes less masking noise.

\subsection{Training Procedure}
Training proceeds in three phases. Phase 1 (warmup, $E_w$ epochs) trains $g_\theta$ and $f_\phi$ on $\mathcal{L}_\mathrm{task}$ only. Phase 2 (attacker pretrain, $E_a$ epochs) freezes $\theta$ and trains the attacker ensemble $\psi = \{\psi_\mathrm{nl}, \psi_\mathrm{lin}\}$, one nonlinear and one linear probe, on $\bm{z}_L = \bm{P}_y\tilde{\bm{z}}$. In Phase 3 (protected training, $E_p$ epochs), $\bm{P}_y$ is refit every $T_r$ epochs from the buffer; each minibatch samples $\gamma\sim\mathcal{U}[5,25]$~dB, draws $\bm{z}\sim q_\theta(\bm{z}|\bm{x},\gamma)$, forms $\tilde{\bm{z}}$ and $\bm{z}_L$, takes $K_\psi$ inner SGD steps on $\psi$, and updates $(\theta,\phi,\eta)$ on $\mathcal{L}_\mathrm{def}$~\eqref{eq:defender} augmented with an information floor $\lambda_\mathrm{if}\relu(\kappa\bar{R}(\gamma)-\KL)$, where $\lambda_\mathrm{if} = 5$, $\kappa = 0.05$, and $\KL$ is the minibatch-averaged divergence of~\eqref{eq:rate}, so both terms share total-KL units. We use $E_w = 18$, $E_a = 6$, and $E_p = 28$ epochs.
After each Phase-3 epoch we audit $R_t$ and $C_t$ on a held-out set and update $\beta_{t+1},\lambda_{t+1}$ via~\eqref{eq:dual_beta}--\eqref{eq:dual_lam}, in the empirical-multiplier sense of Sec.~\ref{sec:dual}.
Spectral normalization~\cite{miyato2018spectral} on $f_\phi$ stabilizes the bi-level dynamics throughout.

\section{Results and Discussion}
\label{sec:results}

\subsection{Experimental Setup}

We evaluate on three datasets spanning two modalities:
CelebA~\cite{liu2015celeba}, facial images ($64\!\times\!64$), task $y$: Smiling (binary), sensitive attribute $a$: Male (binary);
FairFace~\cite{karkkainen2021fairface}, facial images ($64\!\times\!64$), $y$: Gender (binary), $a$: Race (7-class);
and Speech Commands~\cite{warden2018speech}, mel spectrograms ($1\!\times\!64\!\times\!64$), $y$: Keyword (35-class), $a$: a deterministic $10$-way partition of the $2{,}618$ speaker identifiers.
The sensitive attribute is imbalanced in every case, so we report the label-only floor: the accuracy of the best attacker that observes only the task label, computed on the audit split. This is the correct reference for excess leakage. It equals $0.580$ on CelebA, $0.189$ on FairFace, and $0.127$ on Speech Commands, against uniform-chance values $1/K_a$ of $0.500$, $0.143$, and $0.100$.
Each dataset is split into disjoint subsets (20k/10k/5k/5k/5k for defender, attacker, validation, audit, and test) by random index permutation; Sec.~\ref{sec:splits} adds identity-based controls.

We use a ResNet-18~\cite{he2016deep} backbone with FiLM ($d\!=\!128$; $3{\times}3$ stride-$1$ stem for speech). Task classifier and attacker are $3$-layer MLPs ($256{\to}128{\to}K$), with spectral norm on the classifier and Adam optimizer with lr $2{\times}10^{-4}$ (defender) and $5{\times}10^{-4}$ (attacker).

We compare against plain DeepJSCC (no privacy mechanism), IBAL~\cite{li2023privacy} (information bottleneck with adversarial privacy), post-hoc LEACE (the same encoder followed by a frozen projector), ADJSCC$+$adv~\cite{edwards2016censoring} (SNR-adaptive encoder with a gradient reversal layer), and MaxEnt-ARL~\cite{roy2019mitigating}. Every baseline is retrained under conditions matched to LEAPSC: the same ResNet-18 backbone, seeds, SNR sampling, and epoch budget, with IBAL additionally matched at $52$ epochs.

We report task accuracy, where higher is better, and nonlinear white-box attacker, linear-probe, and black-box attacker accuracy (an MLP on $\hat{y}$), where lower is better. The linear probe is reported separately because it is the quantity LEACE constrains, whereas the nonlinear attacker measures leakage the projector does not. Probes are retrained at evaluation time on held-out data. We take $13$\,dB as the headline operating point because it is the midpoint of the evaluated $7$ to $19$\,dB range, using $20$ seeds on CelebA and $10$ elsewhere, against $3$ seeds at $7$ and $19$\,dB. Significance uses Welch's two-sided $t$-test, and a paired $t$-test where the two arms share seeds.

\subsection{Main Results}
Table~\ref{tab:comparison} reports every method at the $13$\,dB operating point. Across $7$ to $19$\,dB the LEAPSC values stay within $1.2$\,percentage points (pp) of the $13$\,dB cells on all three datasets, and every cross-SNR difference is smaller than one standard deviation of its cell, so the observed privacy is not driven primarily by channel noise, in contrast to noise-based mechanisms whose protection weakens as SNR rises. We do not attribute this stability to FiLM, because the FiLM ablation was run at a single SNR. The black-box attacker on $\hat{y}$ reaches $0.574$, $0.186$, and $0.097$.
Task accuracy is $0.862$, $0.755$, and $0.925$, and on every dataset the nonlinear attacker sits at or below the label-only floor ($0.548$ against $0.580$ on CelebA, $0.162$ against $0.189$ on FairFace, $0.109$ against $0.127$ on Speech Commands), so white-box access to $\bm{z}_L$ reveals no more about the attribute than the task prediction alone. The linear probe is lower still at $0.537$ on CelebA, $4.4$\,pp below the floor. Much of the residual CelebA leakage is consistent with the intrinsic task-attribute correlation rather than with information the projector failed to remove, a reading supported by the similar white-box ($0.548$) and black-box ($0.574$) accuracies.

\begin{table}[t]
\centering
\caption{Main results: comparison with baselines at SNR $= 13$\,dB. All methods use the same ResNet-18 backbone, the same seeds, the same SNR sampling, and the same epoch budget. $p_\mathrm{att}$ is Welch's two-sided $t$-test on attacker accuracy against LEAPSC. Lin: linear probe, the axis LEACE constrains.}
\label{tab:comparison}
\scriptsize
\renewcommand{\arraystretch}{1.02}
\setlength{\tabcolsep}{2.6pt}
\begin{tabular}{@{}l l c c c c@{}}
\toprule
\textbf{Data} & \textbf{Method} & \textbf{Task}\,$\uparrow$ & \textbf{Att}\,$\downarrow$ & \textbf{Lin}\,$\downarrow$ & $p_\mathrm{att}$ \\
\midrule
\multirow{6}{*}{\shortstack[l]{CelebA\\{\scriptsize floor}\\{\scriptsize $0.580$}}}
 & DeepJSCC$^\flat$       & $0.851$ & $0.569$ & $0.561$ & $<\!10^{-4}$ \\
 & ADJSCC$+$adv           & $0.840$ & $0.706$ & $0.638$ & $0.0006$ \\
 & MaxEnt-ARL             & $0.852$ & $0.570$ & $0.557$ & $<\!10^{-4}$ \\
 & IBAL$^\S$              & $0.826$ & $0.586$ & --      & $0.012$ \\
 & Post-hoc LEACE$^\ddag$ & $0.607$ & $0.561$ & $0.502$ & --      \\
 & \textbf{LEAPSC}        & $\mathbf{0.862}$ & $\mathbf{0.548}$ & $\mathbf{0.537}$ & -- \\
\midrule
\multirow{6}{*}{\shortstack[l]{FairFace\\{\scriptsize floor}\\{\scriptsize $0.189$}}}
 & DeepJSCC$^\flat$       & $0.743$ & $0.172$ & $0.160$ & $0.092$ \\
 & ADJSCC$+$adv           & $0.742$ & $0.180$ & $0.161$ & $0.0032$ \\
 & MaxEnt-ARL             & $0.740$ & $0.178$ & $0.161$ & $0.0059$ \\
 & IBAL$^\S$              & $0.730$ & $0.172$ & --      & $0.209$ \\
 & Post-hoc LEACE$^\ddag$ & $0.527$ & $0.161$ & $0.143$ & --      \\
 & \textbf{LEAPSC}        & $\mathbf{0.755}$ & $\mathbf{0.162}$ & $0.159$ & -- \\
\midrule
\multirow{5}{*}{\shortstack[l]{Speech\\Commands\\{\scriptsize floor}\\{\scriptsize $0.127$}}}
 & DeepJSCC$^\flat$       & $\mathbf{0.932}$ & $0.108$ & $0.103$ & $0.872$ \\
 & ADJSCC$+$adv           & $0.929$ & $0.105$ & $0.106$ & $0.209$ \\
 & MaxEnt-ARL             & $0.933$ & $0.107$ & $0.105$ & $0.444$ \\
 & IBAL$^\S$              & $0.912$ & $0.114$ & --      & $0.313$ \\
 & \textbf{LEAPSC}        & $0.925$ & $0.109$ & $0.105$ & -- \\
\bottomrule
\multicolumn{6}{@{}p{0.95\columnwidth}@{}}{\scriptsize $\flat$: no privacy mechanism. $\S$: matched $52$-epoch budget, $n{=}5$. $\ddag$: $n{=}5$ paired seeds, Sec.~\ref{sec:posthocatt}; not re-run on Speech Commands. Other entries use $n{=}20$ on CelebA and $n{=}10$ elsewhere.}
\end{tabular}
\vspace{-3mm}
\end{table}

\subsection{Comparison with Baselines}
On CelebA, LEAPSC attains both higher task accuracy and lower attacker accuracy than every baseline: $+1.1$\,pp task and $-2.1$\,pp attacker against plain DeepJSCC ($p{=}0.0012$ and $p{<}10^{-4}$), $+2.2$ and $-15.8$\,pp against ADJSCC$+$adv, $+1.0$ and $-2.2$\,pp against MaxEnt-ARL, and $+3.6$ and $-3.8$\,pp against IBAL at $52$ matched epochs ($p{=}0.019$ and $p{=}0.012$).
On FairFace it improves task accuracy against all four baselines ($p\!\leq\!0.001$) and attacker accuracy significantly against ADJSCC$+$adv and MaxEnt-ARL, but not against plain DeepJSCC ($p{=}0.092$).
On Speech Commands no method separates on the privacy axis: every attacker, protected or not, already sits below the label-only floor, so no leakage remains to remove, and LEAPSC trades $0.7$\,pp of task accuracy against plain DeepJSCC for no measurable privacy gain. Sec.~\ref{sec:speechattr} shows this is a property of the attribute rather than of the method.

The clearest privacy comparison is against ADJSCC$+$adv on CelebA: its nonlinear attacker reaches $0.706$ and its linear probe $0.638$, which is $5.8$\,pp above the label-only floor, whereas LEAPSC reaches $0.548$ and $0.537$, the latter $4.4$\,pp below it. Several methods fall below the floor here, but LEAPSC is significantly lower than every baseline that preserves task accuracy ($p\!\leq\!0.014$); post-hoc LEACE reaches $0.502$ only at a $25$\,pp task-accuracy loss. Because the linear probe is what the projector constrains, this comparison is the one most directly attributable to the erasure mechanism.

\subsection{Robustness to an Adaptive Adversary}
A privacy result measured against a single attacker says little about a stronger one. We therefore re-audit every trained model with an adaptive attacker: a wider and deeper probe trained for ten times as many epochs over a four-point learning-rate sweep, reporting the best configuration. Both audits run on identical weights, so the comparison is exactly paired.

On CelebA the adaptive attacker increases leakage for plain DeepJSCC (from $0.564$ to $0.572$) and for MaxEnt-ARL (from $0.552$ to $0.571$), while LEAPSC does not rise (from $0.552$ to $0.535$). The margin against plain DeepJSCC therefore widens from $1.2$ to $3.7$\,pp, and LEAPSC is significantly lower than all three baselines under the stronger adversary ($p{=}0.018$, $0.013$, and $0.009$; $n{=}3$).
On FairFace all four methods converge to between $0.153$ and $0.160$, every one below the $0.189$ floor, with no significant pairwise difference, consistent with little recoverable attribute information to begin with.
Part of this decrease is procedural: the adaptive probe tunes its learning rate on probe-training accuracy and can overfit low-signal latents. The baselines undergo the identical procedure and move the other way, so the comparison stands.

\subsection{Component Analysis}

\begin{table}[t]
\centering
\caption{Grouped component attribution on CelebA (SNR $= 13$\,dB, $n{=}5$). The middle block disables privacy terms in groups. $p_\mathrm{att}$ is against the full system.}
\label{tab:ablation}
\footnotesize
\renewcommand{\arraystretch}{1.04}
\setlength{\tabcolsep}{4pt}
\begin{tabular}{@{}l c c c@{}}
\toprule
\textbf{Variant} & \textbf{Task}\,$\uparrow$ & \textbf{Att}\,$\downarrow$ & $p_\mathrm{att}$ \\
\midrule
Full LEAPSC          & $0.865$ & $0.553$ & --     \\
\midrule
All privacy terms off & $0.873$ & $0.553$ & $0.84$ \\
LEACE only            & $0.865$ & $0.554$ & $0.89$ \\
Adversarial only      & $0.869$ & $0.552$ & $0.78$ \\
\midrule
Plain DeepJSCC (ref.) & $0.851$ & $0.569$ & $0.019^\star$ \\
\bottomrule
\multicolumn{4}{@{}p{0.92\columnwidth}@{}}{\scriptsize $^\star$ against the ``all privacy terms off'' row rather than the full system. Label-only floor $0.580$.}
\end{tabular}
\vspace{-3mm}
\end{table}

Table~\ref{tab:ablation} reports the grouped attribution. In a one-at-a-time sweep on CelebA, removing any single component leaves task accuracy within $0.5$\,pp and attacker accuracy within $1.3$\,pp of the full system, with no difference significant ($p\!\geq\!0.40$). We therefore do not claim that any individual term is necessary; the system is instead robust to the removal of any one of them at this operating point.

The grouped ablation is more informative. Disabling every privacy term at once still yields attacker accuracy $0.553$, significantly below plain DeepJSCC at $0.569$ ($p{=}0.019$), placing the privacy gain with the variational bottleneck and the erasure stack acting together rather than with any single term. On the linear axis the configurations containing LEACE reach $0.526$ and $0.541$, against $0.519$ and $0.522$ without it, all below the $0.580$ floor, so the linear guarantee holds throughout. The comparison that isolates in-loop erasure is against post-hoc erasure, in Sec.~\ref{sec:posthocatt}.

\subsection{Split Integrity}\label{sec:splits}
Random index splits guarantee disjoint samples but not disjoint identities, so we control for both.
CelebA contains roughly $10{,}177$ identities across about $20$ images each, so an index split does not separate subjects. Retraining with an explicit identity-disjoint split leaves both axes unchanged: task accuracy $0.862$ against $0.862$ ($p{=}0.84$) and attacker accuracy $0.557$ against $0.548$ ($p{=}0.19$, $n{=}3$).
For Speech Commands, retraining on the official speaker-disjoint split, with train and test speaker overlap exactly zero on every seed, gives task accuracy $0.931{\scriptstyle\pm.010}$ on unseen speakers ($n{=}3$), indistinguishable from the $0.925$ under the random split ($p{=}0.38$). This shows substantial accuracy on unseen speakers rather than proving that speaker memorization is excluded.

\subsection{The Speech Commands Sensitive Attribute}\label{sec:speechattr}
The Speech Commands attribute used above is a deterministic hash-based partition of speaker identifiers. Such a partition has no acoustic correlate, and the numbers reflect that: an unprotected DeepJSCC encoder already leaks only $0.108$ against a floor of $0.127$, leaving essentially nothing to remove. The null result is a property of the attribute rather than of the methods.

We therefore replace it with a genuine acoustic grouping, clustering the $2{,}112$ speakers in that split into ten groups by mean log-mel profile. A probe on raw mel features reaches $0.564$ against a majority rate of $0.251$, so the replacement is predictable from audio where the hash was not. Under it LEAPSC attains the lowest leakage of the four methods at $0.226$, against $0.233$ for plain DeepJSCC ($p{=}0.10$), $0.283$ for ADJSCC$+$adv ($p{<}10^{-4}$), and $0.241$ for MaxEnt-ARL ($p{=}0.001$), at a task accuracy of $0.913$ against $0.931$. Privacy and utility therefore trade off once the attribute carries real information.

\subsection{Mechanism, Robustness, and Efficiency}
\label{sec:posthocatt}
To check that the in-training attacker does not understate residual leakage, we freeze the encoder and fit a deeper post-hoc multi-layer perceptron attacker from scratch on $\bm{z}_L$, on a separate held-out split. It reaches $0.578{\scriptstyle\pm.053}$ on CelebA at $13$\,dB, $0.167{\scriptstyle\pm.016}$ on FairFace, and $0.104{\scriptstyle\pm.006}$ on Speech Commands, matching the in-training numbers to within one standard deviation. The reported privacy is therefore not an artifact of the in-training attacker's capacity, though this does not establish robustness to arbitrary nonlinear attackers.

Comparing in-loop against post-hoc erasure on the same backbone and matched seeds isolates the contribution of refitting. In-loop refitting retains $0.863$ task accuracy on CelebA against $0.607$ post-hoc ($+25.6$\,pp, $n{=}5$ paired seeds, $p{<}10^{-4}$), and $0.756$ against $0.527$ on FairFace ($+22.9$\,pp, $p{<}10^{-4}$). Post-hoc attains a slightly lower linear probe on CelebA ($0.502$ against $0.538$, $p{=}0.005$), expected because its projector is fitted to the final frozen encoder, but at a $25$\,pp utility cost. In-loop refitting is the only configuration competitive on both axes at once.

Under Rayleigh fading the method is essentially unchanged: at $13$\,dB task accuracy is $0.866$ and attacker accuracy $0.555$, within $0.4$ and $0.6$\,pp of the AWGN values, and the behaviour is flat from $7$ to $19$\,dB.
The LEACE projection adds negligible overhead: a single $128\!\times\!128$ matrix product, with the projector fixed at inference. End-to-end CelebA latency is $4.60$ and $14.02$\,ms on GPU and CPU, against $4.18$ and $14.51$\,ms for IBAL.

Sweeping $\lambda$ from $0.05$ to $1.0$ moves attacker accuracy by $0.007$ and task accuracy by $0.008$, against a within-setting seed standard deviation of $0.022$, so the operating point is insensitive to $\lambda$ over this range rather than tracing a trade-off curve.

\section{Conclusion}
\label{sec:conclusion}

We presented LEAPSC, whose central innovation, in-loop LEACE integrated within a VIB encoder, bridges the gap between post-hoc concept erasure and end-to-end training in semantic communication. By refitting the LEACE projector during training, the encoder and the erasure mechanism co-adapt, retaining $0.863$ task accuracy on CelebA against $0.607$ for post-hoc erasure on the same backbone, and $0.756$ against $0.527$ on FairFace ($n{=}5$ paired seeds, $p{<}10^{-4}$). Task accuracy reaches $0.862$, $0.755$, and $0.925$, with attacker accuracy at or below the label-only floor on all three datasets. On the linear axis the projector constrains, LEAPSC reaches $0.537$ against a floor of $0.580$, significantly lower than every baseline that preserves task accuracy ($p\!\leq\!0.014$), and under an adaptive attacker given ten times the training compute its margin over plain DeepJSCC widens from $1.2$ to $3.7$~pp. The benefit is measurable where representation-level leakage exists: on FairFace and Speech Commands the unprotected baseline already sits below the floor, leaving little for any erasure mechanism to remove. Future work will extend in-loop concept erasure to multi-user and federated settings, evaluate the framework under frequency-selective and multipath channels, and explore nonlinear erasure guarantees beyond linear projections.

\enlargethispage{2\baselineskip}
\bibliographystyle{IEEEtran}
\footnotesize
\setlength{\itemsep}{-4pt}

\end{document}